\documentclass[aps,prl,twocolumn,amsmath,amssymb,floatfix,showpacs]{revtex4}
\usepackage{graphicx}
\usepackage{dcolumn}
\usepackage{bm}
\usepackage{psfig}

\newcommand{\mbf}[1]{\boldsymbol{#1}}
\newcommand{\dd}{\mathrm{d}}
\begin{document}

  \title{Diffusive transport in spin-1 chains at high temperatures}
  
  \author{J. Karadamoglou}
  \affiliation{Institut Romand de Recherche Num\'erique en
   Physique des Mat\'eriaux (IRRMA), \\ EPFL, 1015 Lausanne,
   Switzerland}

  \author{X. Zotos}
  \affiliation{Department of Physics, University of Crete  and Foundation 
   for Research and Technology-Hellas, P. O. Box 2208, 71003 Heraklion, 
   Crete, Greece}
  
\date{\today}
  \begin{abstract}
    We present a numerical study on the spin and thermal conductivities 
    of the spin-1 Heisenberg chain in the high temperature limit, 
    in particular of the Drude weight contribution and frequency dependence.
    We use the Exact Diagonalization and the recently developed 
    microcanonical Lanczos method; it allows us a finite size 
    scaling analysis by the study of significantly larger lattices.
    This work, pointing to a diffusive rather than ballistic behavior 
    is discussed with respect to other recent theoretical and experimental 
    studies.
  \end{abstract}
  \maketitle

  \emph{Introduction.--} Recently, numerous experiments on quasi-one 
  dimensional (1D)  
  spin-1/2 compounds \cite{ SologubenkoGOAR, SologubenkoGOVR, HessBABMBR,
  KudoINAKMTK, SologubenkoODR} have confirmed highly anisotropic thermal
  transport along the direction of the magnetic chains and a large
  contribution to the thermal conductivity due to the magnetic interactions. 
  This is in agreement with early theoretical 
  proposals \cite{HuberS, NiemeijerV} of ballistic
  transport in spin-1/2 Heisenberg antiferromagnetic chains (HAFM), 
  that was recently related to the integrability of 
  this system \cite{review, ZotosNP, SaitoTM, NarozhnyMA}. These 
  developments promoted the theoretical study of several models, as 
  spin-1/2 frustrated chains, ladders and higher spin systems, using 
  numerical methods \cite{AlvarezG, HeidrichHCB, Zotos} or
  low energy effective theories 
  \cite{RoschA,SachdevD, BuragohainS,OrignacCC, Saito}.

  On the spin-1 compound AgVP$_2$S$_6$ \cite{SologubenkoKO},  
  thermal conductivity experiments revealed 
  anisotropic transport - qualitatively similar to that of 
  spin-1/2 compounds - while NMR \cite{TakigawaAAMU} concluded to 
  diffusive spin transport at high temperatures and suggested a change in 
  behavior at low temperatures. The $S=1$ HAFM model is nonintegrable and 
  its physics characterized by a finite excitation gap \cite{Haldane}. 
  Although there has been significant progress in understanding the 
  thermodynamics of $S=1$ compounds, 
  there are still open questions regarding transport. 
  In particular, theoretical analysis based on a semiclassical approach of the 
  quantum non-linear sigma model (NL$\sigma$M) \cite{SachdevD, BuragohainS}     
  - the standard low energy description of $S=1$ chains and an 
  integrable model -  concluded to diffusive dynamics while  
  a Bethe ansatz method calculation \cite{Fujimoto,Konik} to ballistic 
  transport.

  The present experimental and theoretical status opens two 
  perspectives that motivate this work; first, once the 1D magnetic 
  transport was established as a new mode of thermal conduction, 
  the ongoing synthesis and study of novel 
  compounds demands the theoretical 
  characterization of conductivities - ballistic vs diffusive - 
  in various spin models. Second, the 
  conjectured connection of ballistic (dissipationless) transport 
  to the integrability of systems requires 
  further theoretical analysis and confirmation.
  
  In this paper, we present a numerical analysis of the  
  thermal and spin transport properties of the spin-1 HAFM system
  in an attempt to obtain a first, albeit for
  finite size lattices, exact picture of the 
  finite temperature/frequency dynamics of this prototype model. 
  We focus the analysis to high temperatures in order to minimize 
  finite size effects and draw reliable conclusions on 
  the thermodynamic limit. In particular, 
  we evaluate the thermal/spin Drude weights, used as the
  criterion of ballistic or diffusive transport. 
  Additionally, we perform calculations
  for the spin $\sigma(\omega)$ and thermal $\kappa(\omega)$ 
  conductivity spectra using
  the Exact Diagonalization (ED) and the recently developed 
  Microcanonical Lanczos Method \cite{LongPSKZ,Zotos} (MCLM) which allows 
  us to obtain results for larger systems than hitherto accessible.
  The data can be used as a benchmark in the development of 
  analytical theories and in the interpretation of experiments in spin-1 
  compounds.

\bigskip
  \emph{Model and Method.--} The Hamiltonian of the spin-1 HAFM chain is
  \begin{equation}
    H = J \sum_{l=1}^L \mbf{S}_{l} \cdot \mbf{S}_{l+1},
    \label{eq:Ham}
  \end{equation}
  where $\mbf{S}_{l}$ is a spin-1 operator at site $l$ and $J$ the
  exchange constant. We consider periodic boundary conditions
  and set $J = \hbar = k_B = 1$.  The spin $j^z$ and energy $j^E$ 
  current operators obtained from the continuity equations for the local 
  spin $S^z$ and energy $H$ are,
  \begin{eqnarray}
    j^z &=& J \sum_{l=1}^{L} 
    S_{l}^{x} S_{l+1}^{y} - S_{l}^{y} S_{l+1}^{x} , \\
    j^E &=&
    J^2 \sum_{l=1}^{L}  \sum_P
    (-1)^P \, S_{l-1}^{P_1} S_{l}^{P_2} S_{l+1}^{P_3},
    \label{eq:j}
  \end{eqnarray}
  where $P$ are the permutations of $x,\,y,\,z$.  

  Within linear response theory \cite{Kubo,Luttinger,NaefZ} 
  the real part of the
  thermal conductivity at frequency $\omega$ and temperature $T$ is given by
  \begin{equation}
    \kappa(\omega) = 2 \pi D_{th} \delta(\omega) + 
    \kappa_{reg}(\omega)
    \label{eq:k'(w)},
  \end{equation}
  where the regular part of the conductivity $\kappa_{reg}$ is 
  \begin{equation}
    \kappa_{reg}(\omega) =
    \frac{\beta}{\omega L} 
    \tanh \Big( \frac{\beta \omega}{2} \Big) 
    \Im i \int^{+\infty}_{0} \!\!\!\!\!\! 
    \dd t e^{izt} \langle \{ j^E(t),j^E \} \rangle ,
    \label{eq:k_reg(w)}
  \end{equation}
  and the thermal Drude weight $D_{th}$ is obtained from
  \begin{equation}
    D_{th} = \frac{\beta^2}{2L} \sum_{\stackrel{n,m}{\epsilon_n=\epsilon_m}} 
    p_n |\langle m | j^E | n \rangle |^2.
    \label{eq:D_th}
  \end{equation}
  Here $\beta = 1 / T$, $z = \omega +i \eta$, $p_n$ are the Boltzmann
  weights and $|n \rangle$ ($|\epsilon_n \rangle$) 
  the eigenstates (eigenvalues), while in Eq.
  (\ref{eq:k_reg(w)}) the symbol $\langle \rangle$ denotes a thermal
  average. In the $\beta \rightarrow 0$ limit we can derive the sum-rule
  \begin{equation}
    \int_{-\infty}^{+\infty} \dd \omega \kappa(\omega) = 
               \frac{\pi \beta^2}{L} \langle (j^E)^2 \rangle = I,
    \label{eq:Sum}
  \end{equation}
  suggesting that a measure of the ballistic contribution to the 
  conductivity is given by the quantity $2\pi D_{th}/I$.
  The corresponding equations for the regular part of the spin conductivity
  $\sigma_{reg}(\omega)$ and Drude weight $D$, can be obtained from
  Eqs. (\ref{eq:k_reg(w)},\ref{eq:D_th}) above by replacing $j^E$ by $j^z$ 
  and dividing them by $\beta$.

  Drude weight data are obtained by using ED which
  restricts us to system sizes up to $L=12$ sites. We use the translational
  and spin symmetries of our system to perform the calculation in
  subspaces of momentum $k$ and magnetization
  $S^z_{tot}$. We find that the results obtained in the $k=0$, $S^z_{tot}=0$
  subspace for $L=12$ (space dimension $\simeq 6500$) 
  are very close to those obtained by diagonalizing the entire 
  Hilbert space.

  For the high temperature $\kappa(\omega)$ and $\sigma(\omega)$ calculations
  we employ the MCLM method \cite{LongPSKZ,Zotos} which allows us to obtain
  results for systems up to $L=18$ sites.  The spectra calculated using this
  method include the Drude weight as a low frequency peak with width of the
  order of the frequency resolution of the method; notice however that this 
  contribution is negligible for the larger systems we 
  study as it follows from the finite size scaling of the Drude weights 
  (see Figs. 1,4). Here we have used $\sim
  1000$ Lanczos steps for the first Lanczos procedure and $\sim 4000$ Lanczos
  steps for the continuous fraction expansion which results in an $\omega$
  resolution of $\sim 0.01$.

\begin{figure}
    \includegraphics[width=0.4\textwidth]{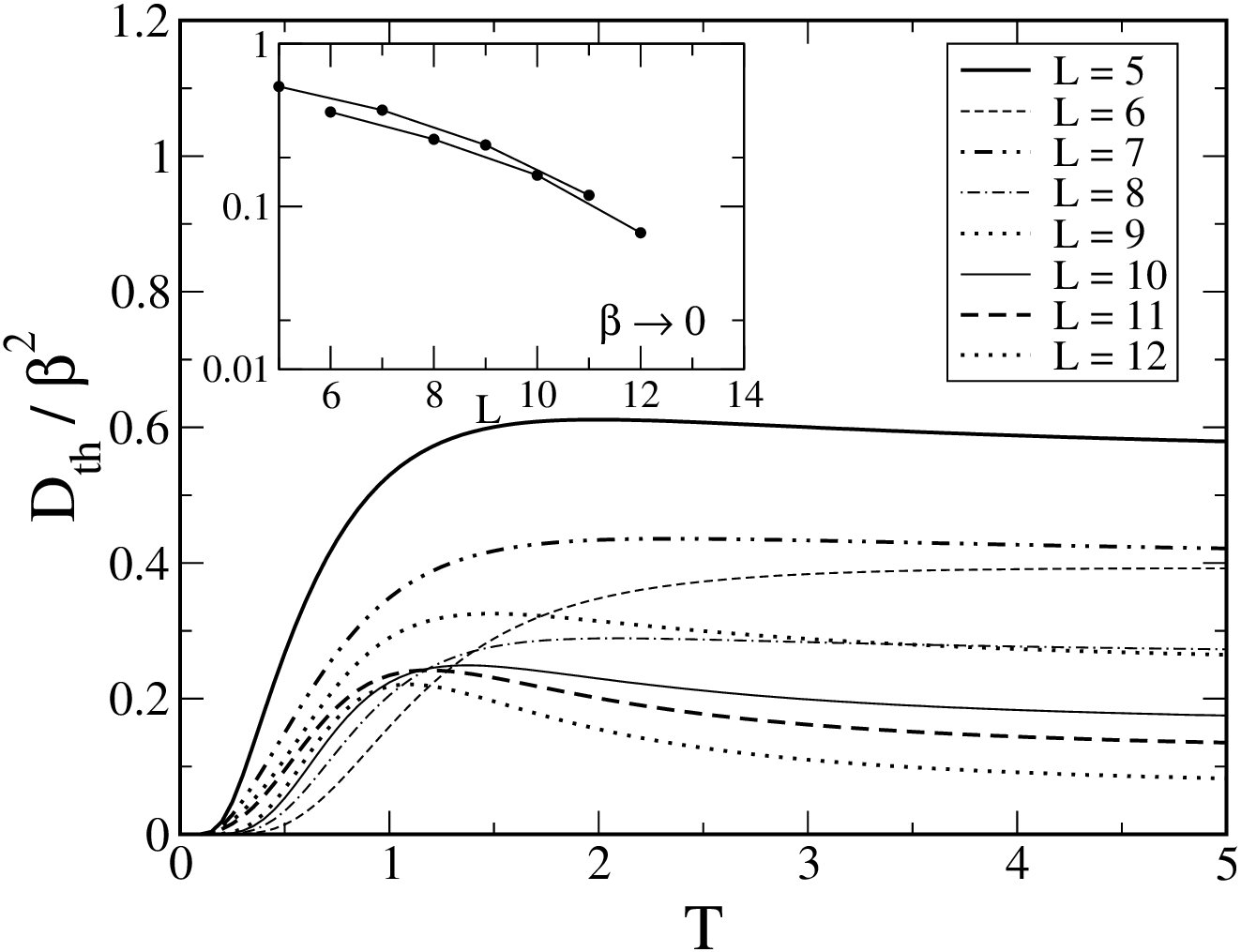}
    \caption{ The thermal Drude weight as a function of temperature for
    several system sizes. In the inset, the scaling of
    $D_{th}/\beta^2$ for $\beta \rightarrow 0$ for odd and even number chains.}
    \label{fig:EScale}
  \end{figure}

\bigskip
  \emph{Thermal conductivity.--} 
  In Fig. \ref{fig:EScale} we show the temperature
  dependence of the thermal Drude weight for several system sizes $L$.
  $D_{th}$ is vanishing at $T=0$ while at high temperatures it has
  a simple $\beta^2$ dependence.  A nonzero Drude weight 
  is generally expected for systems with size less than the
  mean free path of the magnetic excitations. 
  In the $\beta\rightarrow 0$ limit, as $L$ increases, 
  $D_{th}/\beta^2$ decreases - seemingly exponentially fast - 
  and appears to scale to zero in the thermodynamic limit as seen by the 
  curves in the inset of Fig. \ref{fig:EScale} (for chains with even and 
  odd number of sites). Our data therefore suggest diffusive thermal 
transport for the spin-1 HAFM chain.

\begin{figure}
    \includegraphics[width=0.4\textwidth]{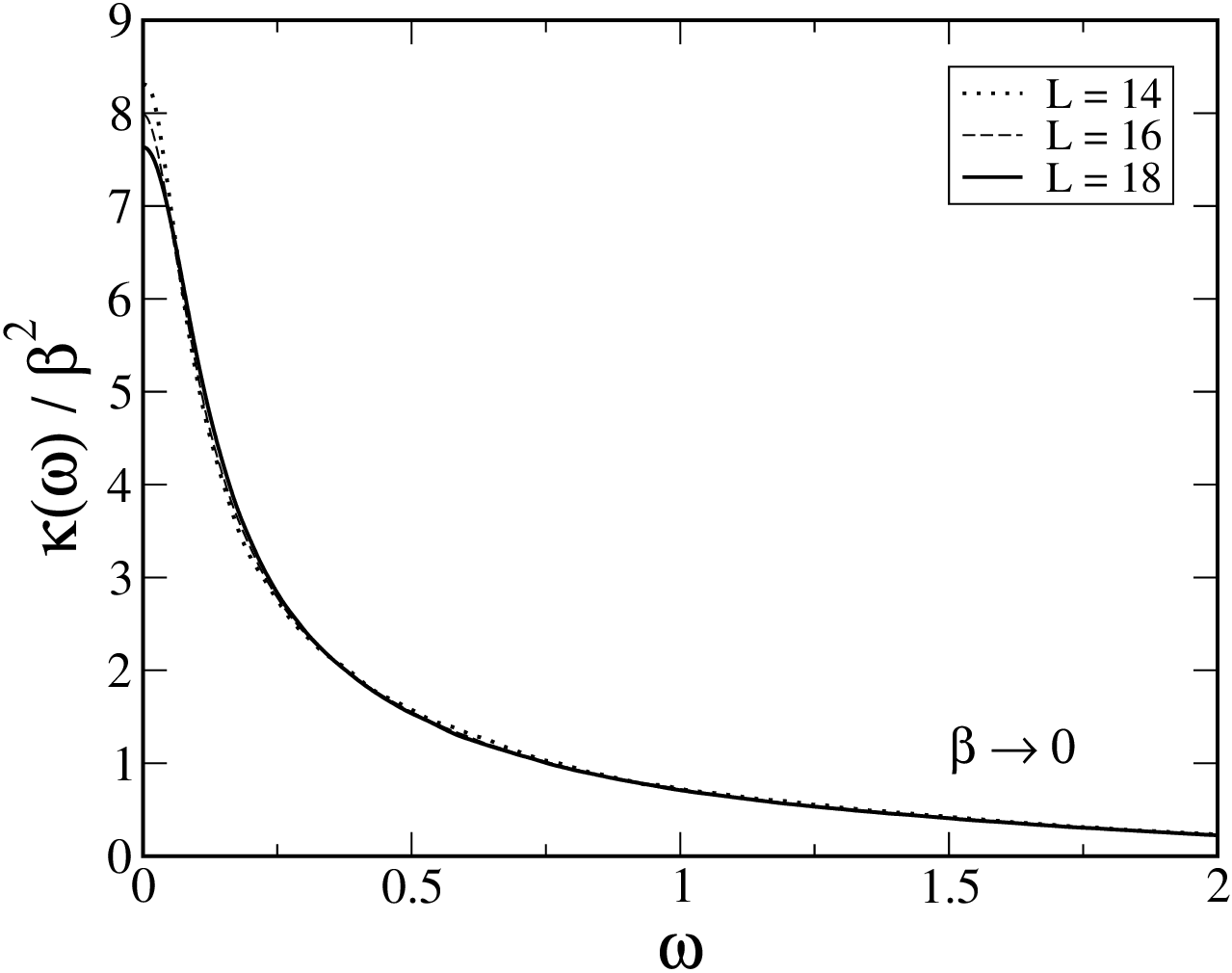}
    \caption{ The thermal conductivity $\kappa(\omega)$
    for $\beta \rightarrow 0$.}
    \label{fig:Kappa}
  \end{figure}

  We now apply the MCLM method to calculate the $\omega$ dependence of
  the thermal conductivity $\kappa(\omega)$ in the high temperature
  limit as shown in Fig. \ref{fig:Kappa}.  For frequencies $\omega
  \gtrsim 0.05$ the conductivity obtained for $L=18$ has practically 
  converged to the $L \rightarrow \infty$ limit while in the
  low frequency regime there is a remaining size dependence. This is
  partly due to the variation of the Drude weight which contributes to
  the low frequency $\kappa(\omega)$.  It is worth noting that the
  statistical fluctuations in our MCLM results are very small, even for
  the smallest size system displayed here. A comparison of ED versus
  MCLM results for $L=12$ (not shown for clarity) gives satisfactory 
  agreement although for
  this size system the statistical fluctuations are significant.

  On the low frequency region, it is not well described by a Lorentzian, as
  predicted by the diffusion phenomenology \cite{KadanoffM}, but the overall
  form of $\kappa(\omega)$ is similar to that 
  found in the $S=1/2$ ladder model \cite{Zotos} and other low dimensional
  models \cite{Prelovsek}; it suggests that this may be a generic 
  behavior of conductivity spectra in such systems.
  From this curve we can also extract an estimate \cite{Zotos} of the high 
  temperature 
  $\kappa_{dc}= \kappa(\omega\rightarrow 0)$ thermal conductivity,    
  $\kappa_{dc}\simeq 16 (\beta J)^2 \frac{W}{mk}$,   
  assuming typical lattice constants $O(10 \AA)$ and $J\sim O(1000$ K).  

  For comparison we note that, (i) a $\kappa_{dc} \sim O(1\frac{W}{mK})$
  was observed at temperatures below the gap - $T\sim 0.2J$ - in the 
  compound AgVP$_2$S$_6$ \cite{SologubenkoKO} and (ii) 
  our high temperature $\kappa_{dc}$ 
  for the spin-1 model is an order of magnitude larger than that of the 
  ladder \cite{Zotos}; notice that 
  the spin-1 HAFM has a similar low energy excitation spectrum 
  and thus low temperature behavior as the spin-1/2 two-leg 
  ladder for $J_{\perp} / J \simeq 0.9$. 

\begin{figure}
    \includegraphics[width=0.4\textwidth]
{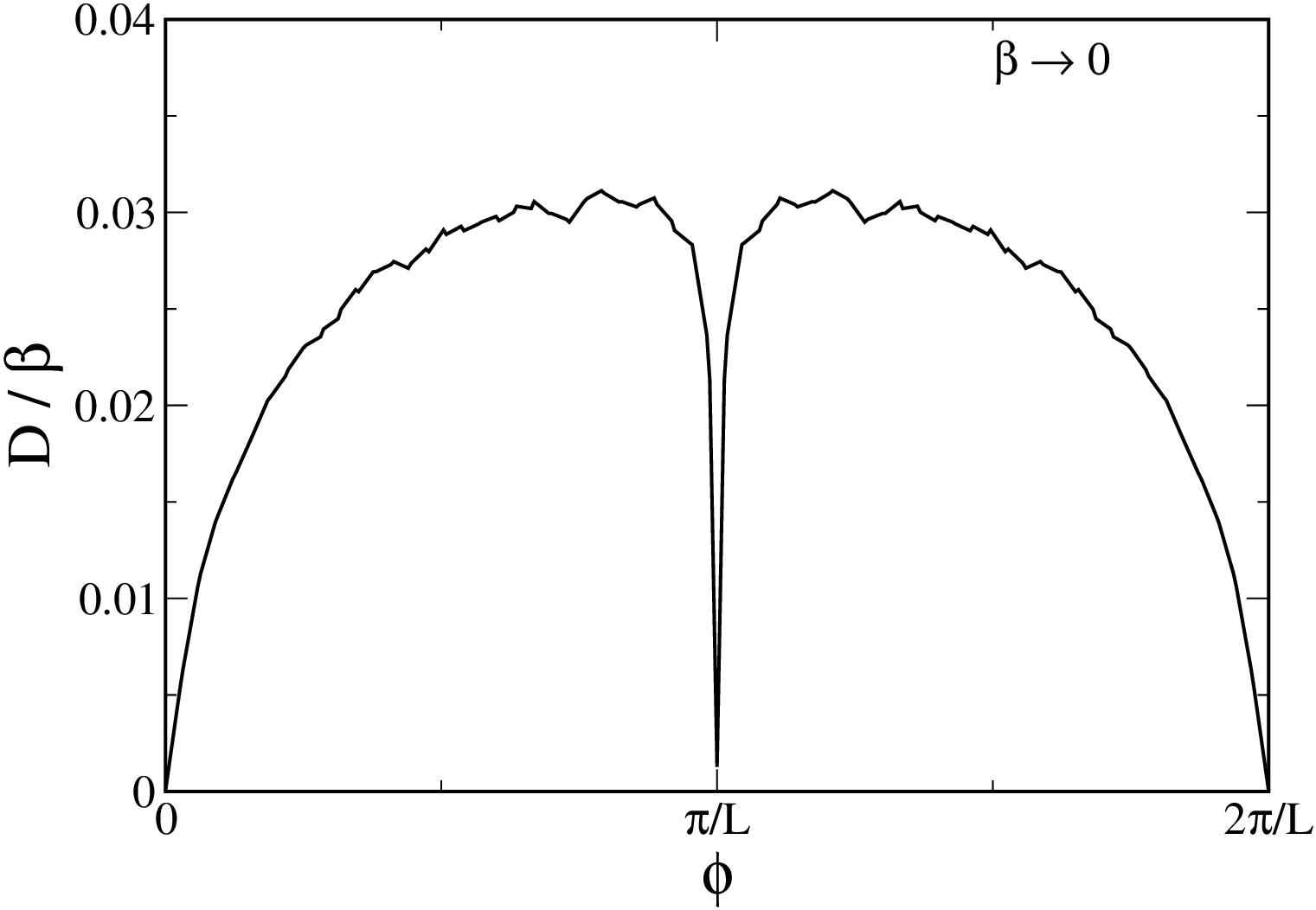}
    \caption{ The $\phi$ dependence of the Drude weight for $L=10$.}
    \label{fig:AllPhi}
  \end{figure}

\bigskip
  \emph{Spin conductivity.--} We now investigate the spin transport by
  calculating the Drude weight $D$ and spin conductivity
  $\sigma(\omega)$.  For $L \leqslant 12$, $D/\beta$ appears to be 
  equal to zero (up to numerical precision) at all temperatures. On this 
  issue it is important to point out that, for faster convergence,  
  we consider only the $S^z=0$ subsector  that is the dominant one 
  in the thermodynamic limit. 
  In order to explore the robustness of this result we apply the canonical
  transformation $S_{l}^{+} \rightarrow S_{l}^{+} e^{i\phi l}$ 
  on $H$ and $j^z$ (periodic in $\phi$ with period $2\pi/L$);  
  the results for $D$ as a function of $\phi$, 
  are shown in Fig. \ref{fig:AllPhi} for $L=10$.  

\begin{figure}
  \includegraphics[width=0.4\textwidth]
  {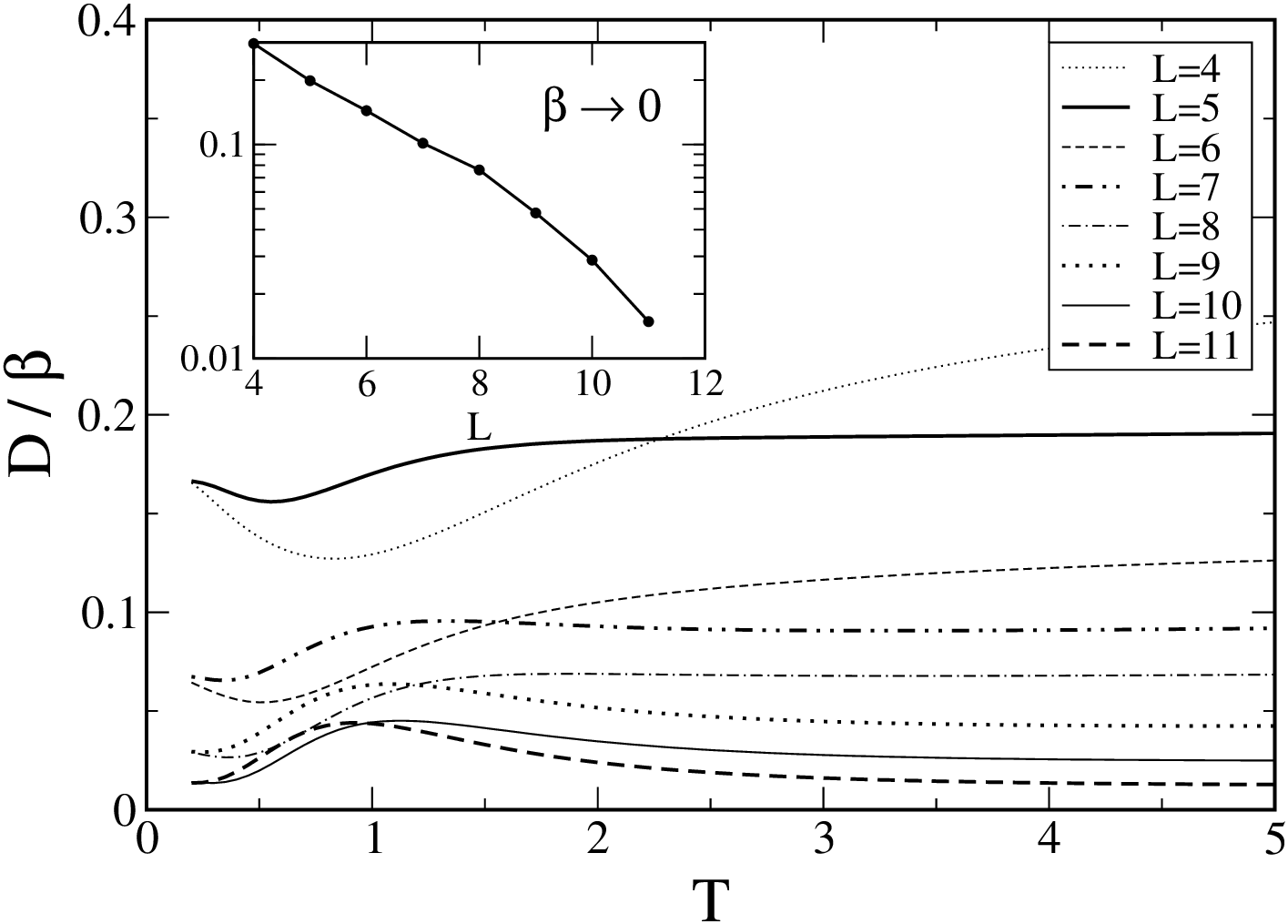}
    \caption{ Drude weight $D$ as a function of 
    temperature for several system sizes $L$. In the inset, the scaling 
    of $D/\beta$ for $\beta\rightarrow 0$.}
    \label{fig:Scale}
  \end{figure}

  We find that $D$ is finite for all $\phi$, except for $\phi=0$ and 
  $\phi=\pi/L$ where it develops a sharp minimum.  
  Curves for different $L'$s show very similar $\phi$
  dependence, but with the local minimum at $\phi=\pi/L$ becoming
  sharper with increasing $L$. We therefore conclude that the vanishing 
  Drude weight at $\phi=0$, even for small $L$, is an artifact of the 
  periodic boundary 
  conditions; notice that a vanishing $D/\beta$ and a nontrivial $\phi$ 
  dependence is also found in the $S=1/2$ isotropic model. 
  In Fig. \ref{fig:Scale} we show that $D/\beta$, close to its maximum value 
  at $\phi=\pi/2L$, is finite throughout the temperature range and
  as shown in the inset, it scales to zero exponentially fast with $L$ in the 
  $\beta\rightarrow 0$ limit. In contrast to the thermal Drude weight, 
  $D/\beta$ goes to a finite value at very low
  temperatures that can be understood considering that a $\phi$ implies 
  a ground state carrying a nonzero spin current. 

\begin{figure}
  \includegraphics[width=0.4\textwidth]{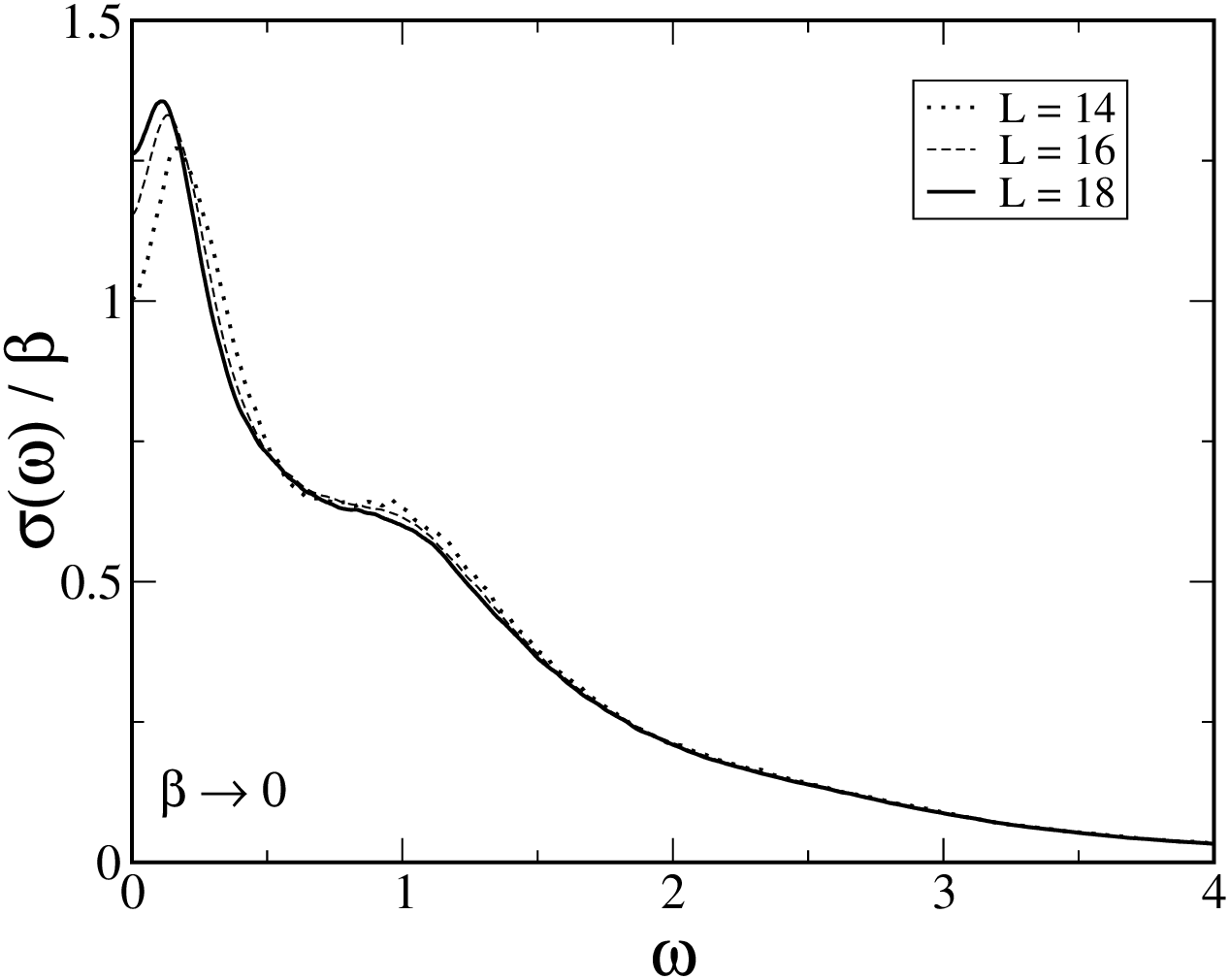}
  \caption{ The spin conductivity $\sigma(\omega)$
  for $\beta \rightarrow 0$. }
  \label{fig:Sigma}
  \end{figure}

  Finally, we present in Fig. \ref{fig:Sigma} $\sigma(\omega)$  
  evaluated using the MCLM method.  We see that there are some 
  statistical fluctuations in the data for the smaller systems while 
  those for the larger systems are very smooth. The curves seem 
  converged to their $L \rightarrow \infty $ limit for $\omega
  \gtrsim 0.2 $.  The main characteristics of our results is the appearance of
  a local maximum at $\omega \sim 1$ and a minimum at $\omega \lesssim 0.2$.
  The later disappears with increasing system size while again we see
  no signs of a Drude peak in the $\sigma(\omega)$ curves.
  It is interesting to note that the local maximum feature has also 
  previously been observed in a study of correlations 
  of the NL$\sigma$M \cite{BuragohainS}.

\bigskip
 \emph{Discussion.--} The overall picture emerging from the presented
  numerical data shows that the high
  temperature spin and energy transport of the spin-1 HAFM chain is
  characterized by finite {\it dc} values, vanishing Drude weights, 
  a smooth frequency dependence (though not of a 
  Lorentzian form) and thus non-ballistic character. This behavior is
  compatible with the assumption of normal transport in nonintegrable 
  models, it is qualitatively 
  similar to that of spin-1/2 antiferromagnetic ladder and 
  in contrast to the ballistic transport of the integrable spin-1/2 version.
  On this point we should mention that 
  in the {\it isotropic} spin-1/2 model $D$ also seems to vanish \cite{review}; 
  however $\sigma(\omega\rightarrow 0)$ might diverge \cite{LongPSKZ} 
  and in any case, in the easy-plane anisotropic $S=1/2$ model $D$ is finite, 
  (in contrast to preliminary results on the anisotropic $S=1$ model).
  On the other hand, for $S=1/2$, $D_{th}$ is clearly finite, as the energy 
  current operator commutes with the Hamiltonian, 
  again in contrast to the $S=1$ case. 
  
From our data in Figs. \ref{fig:Kappa} and 
\ref{fig:Sigma} we can also extract the spin 
${\cal D}_{s}=\sigma_{dc}/\chi\sim 1.4 \beta /\frac{2}{3}\beta \sim 2.1$ 
and thermal 
${\cal D}_{th}=\kappa_{dc}/ C\sim 7.5 \beta^2 /\frac{4}{3}\beta^2 \sim 5.6$ 
diffusion constant (in units of $J/\hbar$), where $\chi$ is the static 
susceptibility and $C$ the specific heat. 
In comparison, a standard 
$\beta\rightarrow 0$ moment analysis \cite{HuberS} gives, 
${\cal D}_s=\sqrt {2\pi {S(S+1)}/{3}}\sim 2.1$ and 
${\cal D}_{th}=\sqrt {{\pi S(S+1)}/{3}} /({1-3/4S(S+1))} \sim 2.3$;
the agreement for ${\cal D}_s$ is excellent (also probably fortuitous 
considering the quantum character of the $S=1$ system) while an enhanced 
value is found for ${\cal D}_{th}$.

  Regarding the low temperature behavior, the limitation of our calculation 
  to small size systems (thus a sparse low energy spectrum) does not allow 
  us to make any reliable statements and in particular, to study an 
  eventual change of transport from diffusive to ballistic as 
  suggested by the experimental results for AgVP$_2$S$_6$
  \cite{TakigawaAAMU}. 
  Yet, there exist spin-1 compounds known,  
  with weak values of $J$, for which these data are directly 
  relevant in the interpretation of transport experiments. A crucial 
  issue remains however for future studies, namely the disentanglement 
  of the spin-phonon 
  from the intrinsic spin-spin scattering contribution to diffusion.

  Finally, on the low energy NL$\sigma$M approach \cite{Fujimoto,
  Konik, SachdevD}, this high temperature study
  cannot shed light on the issue of ballistic vs. diffusive behavior.
  If it is concluded that the NL$\sigma$M predicts 
  diffusive transport then there is continuity with the present 
  $\beta\rightarrow 0$ data. If, on the other hand, 
  ballistic transport (perhaps due to the integrability of the 
  NL$\sigma$M) is found then, the omitted ``irrelevant" terms 
  (for the thermodynamics) could result to a diffusive behavior at all 
  temperature scales.
  
\bigskip
\emph{Acknowledgments. --}
J.K. acknowledges financial
support by the Swiss National Science Foundation,  
the University of Fribourg and the 
Foundation for Research and Technology-Hellas.

\end{document}